\documentclass[conference]{IEEEtran}
\IEEEoverridecommandlockouts

\usepackage[utf8]{inputenc}
\usepackage[T1]{fontenc}
\usepackage{cite}
\usepackage{amsmath,amssymb,amsfonts}
\usepackage{graphicx}
\usepackage{url}
\usepackage[hidelinks]{hyperref}
\usepackage{array}
\usepackage{booktabs}
\usepackage{tabularx}
\usepackage{tikz}
\usetikzlibrary{arrows.meta,positioning,calc,shapes.geometric,shapes.symbols,fit,backgrounds,patterns}
\usepackage{xcolor}

\definecolor{rsrc}{HTML}{59A14F}
\definecolor{ring}{HTML}{4E79A7}
\definecolor{rsvc}{HTML}{B5413A}
\definecolor{ranl}{HTML}{8FB17F}
\definecolor{rstr}{HTML}{C6C6C6}
\definecolor{rviz}{HTML}{59B5C9}
\definecolor{rrep}{HTML}{E2862C}
\definecolor{rblk}{HTML}{B49FCE}

\begin{document}

\title{RESILAGENT: A Three-Pillar Architecture for Cyber-Physical
Resilience in Electrical Power and Energy Systems}

\author{%
\IEEEauthorblockN{Nikolaos Kekatos\IEEEauthorrefmark{1},
Georgios Koutidis\IEEEauthorrefmark{1},
Konstantinos Antonakopoulos\IEEEauthorrefmark{2},
Stylianos Basagiannis\IEEEauthorrefmark{3},\\
Sokratis Katsikas\IEEEauthorrefmark{4},
Georgios Kavallieratos\IEEEauthorrefmark{4},
Alexios Lekidis\IEEEauthorrefmark{5},
Tom Nianios\IEEEauthorrefmark{1},
Elpiniki Papageorgiou\IEEEauthorrefmark{5}}
\IEEEauthorblockA{\parbox{0.93\textwidth}{\centering
\IEEEauthorrefmark{1}Clone Systems, Cyprus
~--~
\texttt{\{nkekatos,\,gkoutidis,\,tnianios\}@clone-systems.com}\quad
\IEEEauthorrefmark{2}BITREZUS, Greece
~--~
\texttt{ka@bitrezus.com}\quad
\IEEEauthorrefmark{3}IHU, Greece
~--~
\texttt{basagiannis@ihu.gr}\quad
\IEEEauthorrefmark{4}NTNU, Norway
~--~
\texttt{\{sokratis.katsikas,\,georgios.kavallieratos\}@ntnu.no}\quad
\IEEEauthorrefmark{5}UTH, Greece
~--~
\texttt{\{alekidis,\,elpinikipapageorgiou\}@uth.gr}
}}}

\maketitle
{\renewcommand{\thefootnote}{}\footnotetext{\copyright~2026 IEEE. Personal use of this material is permitted. Permission from IEEE must be obtained for all other uses, in any current or future media, including reprinting/republishing this material for advertising or promotional purposes, creating new collective works, for resale or redistribution to servers or lists, or reuse of any copyrighted component of this work in other works. This work has been accepted for publication at the 2026 IEEE International Conference on Cyber Security and Resilience (IEEE CSR 2026).}}

\begin{abstract}
Electrical Power and Energy Systems face \emph{hybrid threats} that
combine cyber intrusions, physical sabotage and climate-driven
stress, and routinely outpace today's siloed monitoring stacks. We
present RESILAGENT, a concept architecture organised around three
operational pillars, Sensing \& Data Intelligence, Resilience
Analytics \& Reporting, and Incident Response \& Recovery,
mediated by a Cloud-based Orchestrator and exercised through a
closed Detect-Simulate-Decide-Sign loop. The framework fuses agentic
artificial intelligence and multimodal large language models for
cross-modal detection, real-time digital twins for what-if simulation,
and a permissioned blockchain anchor for tamper-evident threat
intelligence and CER/NIS2-aligned evidence. We instantiate it on two
scenarios, a transmission operator under coordinated cyber compromise,
sabotage and heatwave, and a distribution operator under a
supply-chain compromise during a storm, and report an IEEE-bus
micro-benchmark that quantifies each pillar through
false-data-injection attack detection, cyber--physical fusion lift,
N-1 contingency screening and anchor cost.
\end{abstract}


\section{Introduction}\label{sec:intro}

\IEEEPARstart{T}{he} cyber-physical exposure of Electrical Power and
Energy Systems (EPES) is widening. The European Union Agency for
Cybersecurity again ranks energy among the most targeted Union
verticals~\cite{enisa2025,enisa_nis360}, and a decade of incidents,
BlackEnergy / Industroyer against the Ukrainian grid (2015--2016),
Industroyer2 (2022), Colonial Pipeline (2021), the Viasat KA-SAT wiper
(2022), and growing supply-chain advisories on solar inverters and
Distributed Energy Resource (DER) firmware~\cite{enisa2024}, confirms
the threat is no longer hypothetical, while heatwaves, storms and
floods overlap with it to produce \emph{compound disruptions} beyond
any single tool. Yet operators still run siloed stacks: commercial
Security Information and Event Management (SIEM) and Intrusion
Detection System (IDS) appliances for cyber, Closed-Circuit Television
(CCTV) and access control for physical security, hydro-meteorological
feeds for environment, and manual standard-operating-procedure
escalation. Cyber alerts never reach the dispatch console, weather
warnings never reach the Security Operations Centre (SOC), and
Critical Entities Resilience (CER, Directive (EU) 2022/2557)~\cite{cer2022}
and Network and Information Security (NIS2, Directive (EU)
2022/2555)~\cite{nis2} evidence is still curated in spreadsheets,
widening the Mean-Time-To-Detect / Mean-Time-To-Respond (MTTD/MTTR)
gap exactly when the attacker compounds the incident.

\textbf{Contributions.}
This position paper introduces RESILAGENT, a staged three-pillar
architecture for EPES resilience that integrates multimodal agentic
detection, Digital-Twin (DT) contingency analysis and tamper-evident
evidence generation, and validates the minimum viable stack through
an IEEE-bus micro-benchmark. \emph{No closed-loop grid
control originates from a Large Language Model (LLM) in this design:
high-impact actions require explicit operator confirmation and pass a
DT plausibility filter first.} The specific contributions are:
\begin{enumerate}
  \item A \emph{three-pillar architecture}, Sensing \& Data
    Intelligence, Resilience Analytics \& Reporting, Incident
    Response \& Recovery, mediated by a Cloud-based Orchestrator
    and tied to an eight-stage closed-loop workflow.
  \item A \emph{technology-coherent design} in which agentic
    Artificial Intelligence (AI)/multimodal LLMs serve detection, Digital Twins serve
    simulation/decision, and a permissioned blockchain anchors every
    artefact for CER/NIS2 evidence by construction.
  \item An \emph{EPES use-case instantiation} on two parallel
    EU pilots, a national Transmission System Operator (TSO) under
    coordinated cyber \& sabotage \& heatwave, and a Distribution
    System Operator (DSO) under DER firmware compromise \& storm,
    with explicit per-step technology mapping.
  \item An \emph{IEEE-bus micro-benchmark} that quantifies the
    contribution of each pillar on the IEEE 14/30/118-bus cases:
    physics-aware False Data Injection Attack (FDIA) detection, cyber--physical fusion lift,
    DT-driven $N{-}1$ contingency screening, and blockchain anchor
    cost.
\end{enumerate}

Section~\ref{sec:related} surveys the state of the art;
Section~\ref{sec:arch} details the RESILAGENT three-pillar
architecture and the closed-loop workflow;
Section~\ref{sec:scenarios} presents the TSO and DSO application
scenarios; Section~\ref{sec:eval} reports the evaluation roadmap and
the initial micro-benchmark; Section~\ref{sec:challenges} analyses
open challenges; Section~\ref{sec:conclusion} concludes.

\section{State of the Art}\label{sec:related}

\textit{AI-powered detection in EPES.} Classical multivariate IDS for
smart grids, ARIES~\cite{aries2020} and SPEAR SIEM~\cite{spear2021},
outperform commercial appliances but are offline and single-modality;
the 2024--2025 wave added graph-attention / Kolmogorov--Arnold
detectors on Supervisory Control and Data Acquisition (SCADA)
topology~\cite{wu2025graphkan}, LLM-based relay detectors above
97\,\%~\cite{saber2026}, and explainable LLM frameworks for Automatic
Generation Control anomalies~\cite{sharshar2025,zaboli2024}; recent work
also targets best-practice cyber training for power-grid
operators~\cite{kavallieratos2024},
but
operationalising any at fleet scale remains open~\cite{enisa_nis360}.
\textit{Agentic AI and digital twins.} Agentic AI, autonomous
LLM-backed agents with tools, memory and planning, has entered CISA
practitioner guidance on careful adoption~\cite{cisa2025_agentic} and
secure AI in operational technology~\cite{cisa2025_otai}, with
Farid~\cite{farid2015} grounding decentralised post-disruption
reconfiguration; for simulation, Eckhart and Ekelhart~\cite{eckhart2019}
survey DTs as a Cyber-Physical-System (CPS) security primitive and Saad
\emph{et al.}~\cite{saad2020} build an IoT-based DT for microgrid
resilience, yet none expose the DT to an \emph{agent layer} issuing
what-if queries during a live hybrid event, the coupling RESILAGENT
targets. \textit{Blockchain CTI and data spaces.} CTI-sharing surveys
flag integrity, attribution and trust as the persistent
gaps~\cite{tounsi2018,mdpi_bccti2024,spear2021}, motivating
blockchain-anchored designs and the International Data Spaces energy
framework~\cite{idsa_energy2024} that positions Gaia-X / IDS connectors
as NIS2-aligned CTI infrastructure, while production still relies on
best-effort MISP exchanges that miss the CER/NIS2 audit trail.
\textit{EU R\&D lineage.} Despite a decade of EU R\&D across smart-grid
security
(\href{https://cordis.europa.eu/project/id/833955}{SDN-microSENSE},
\href{https://cordis.europa.eu/project/id/101021936}{ELECTRON},
\href{https://cordis.europa.eu/project/id/101020560}{CyberSEAS}),
power-grid resilience and data spaces
(\href{https://cordis.europa.eu/project/id/101075665}{eFORT},
\href{https://cordis.europa.eu/project/id/101075714}{R2D2},
\href{https://cordis.europa.eu/project/id/101069831}{ENERSHARE}),
cyber-physical CI resilience
(\href{https://cordis.europa.eu/project/id/101021668}{PRECINCT},
\href{https://cordis.europa.eu/project/id/101073909}{ATLANTIS}) and
CI cybersecurity with AI
(\href{https://cordis.europa.eu/project/id/101070455}{DYNABIC},
\href{https://cordis.europa.eu/project/id/101070450}{AI4CYBER}), no
operational architecture closes the loop between agentic detection,
digital-twin analysis and verifiable evidence generation while
preserving CER/NIS2 compliance and human approval for high-impact
actions, the gap RESILAGENT targets.

\subsection*{Positioning of the Contribution}
\label{ssec:positioning}
Table~\ref{tab:positioning} positions RESILAGENT against the closest
2024--2026 comparables across seven coverage dimensions: none of them
simultaneously covers all seven, the gap RESILAGENT targets as an
advanced proof-of-concept exercised end-to-end
(Section~\ref{sec:eval}) on two EPES scenarios
(Section~\ref{sec:scenarios}).

\begin{table}[t]
\centering
\caption{Positioning vs.\ comparable frameworks (2024--2026):
(a)~multimodal detection; (b)~real-time DT; (c)~agentic AI/LLM;
(d)~blockchain audit/CTI; (e)~federated learning; (f)~EPES pilot;
(g)~open-source toolchain. $\checkmark$ supported,
$\sim$ partial, $\square$ not present.}
\label{tab:positioning}
\scriptsize
\renewcommand{\arraystretch}{1.0}
\setlength{\tabcolsep}{3pt}
\begin{tabularx}{\columnwidth}{X c c c c c c c}
\toprule
\textbf{Framework} & (a) & (b) & (c) & (d) & (e) & (f) & (g) \\
\midrule
\href{https://cordis.europa.eu/project/id/101075665}{eFORT}    & $\checkmark$ & $\checkmark$ & $\square$    & $\checkmark$ & $\square$    & $\checkmark$ & $\square$    \\
\href{https://cordis.europa.eu/project/id/101075714}{R2D2}    & $\checkmark$ & $\sim$       & $\square$    & $\square$    & $\square$    & $\checkmark$ & $\square$    \\
\href{https://cordis.europa.eu/project/id/101073909}{ATLANTIS}& $\checkmark$ & $\square$    & $\sim$       & $\square$    & $\square$    & $\checkmark$ & $\sim$       \\
Grid-Agent~\cite{gridagent2025}             & $\sim$       & $\square$    & $\checkmark$ & $\square$    & $\square$    & sim          & $\square$    \\
Blockchain CTI~\cite{tounsi2018,mdpi_bccti2024}  & $\square$    & $\square$    & $\square$    & $\checkmark$ & $\square$    & $\square$    & $\square$    \\
MS Defender \& Sentinel~\cite{msdefender2025}& $\checkmark$ & $\square$    & $\sim$       & $\square$    & $\square$    & $\checkmark$ & $\square$    \\
\midrule
\textbf{RESILAGENT (this paper)}            & $\checkmark$ & $\checkmark$ & $\checkmark$ & $\checkmark$ & $\sim$       & $\sim$       & $\checkmark$ \\
\bottomrule
\end{tabularx}
\end{table}

\section{RESILAGENT Three-Pillar Architecture}\label{sec:arch}

We use three terms precisely. An \textbf{agent} is an LLM-backed
component with tools, memory and a planner; modality-specific agents
emit anomalies and a multimodal \emph{fusion agent} consolidates them
into one hybrid-event hypothesis driving a Security Orchestration,
Automation and Response (SOAR) playbook under a human-in-the-loop
checkpoint. A \textbf{real-time Digital Twin} is a continuously
synchronised simulation of grid topology, loads, generation and
protection settings, exposed as a DT-as-a-Service (DTaaS) endpoint and
coupled to cyber emulators via co-simulation
(HELICS~\cite{helics2017}, OPAL-RT Hardware-in-the-Loop, HIL); it
is at once the ``normal'' reference frame, a forward-prediction
engine and a counterfactual sandbox~\cite{bensalem2023}. A \textbf{blockchain anchor}
is a permissioned ledger (Hyperledger Fabric) binding
every artefact, detector output, DT contingency, SOAR action
and Cyber Threat Intelligence (CTI) bundle to a
tamper-evident hash chain; it has no detection role, contributing
verifiability, supply-chain attribution and CER/NIS2 evidence by
construction.

\begin{figure}[t]
\centering
\resizebox{0.84\columnwidth}{!}{%
\begin{tikzpicture}[
  font=\footnotesize,
  pillar/.style={rounded corners=3pt,draw=black!30,
                 inner sep=4pt,align=center,minimum height=1.9cm,
                 minimum width=6.4cm},
  hub/.style={draw=rblk!80!black,fill=rblk!25,rounded corners=4pt,
              thick,minimum width=7.6cm,minimum height=1.0cm,
              align=center,font=\bfseries\small},
  arr/.style={-{Latex[length=2.2mm]},thick,black!65},
  outer/.style={draw=black,line width=1.4pt,rounded corners=10pt,
                inner sep=10pt,inner xsep=14pt},
]
\node[pillar,fill=ring!18] (p1) at (0,0) {
  \begin{minipage}{6.0cm}\centering
  \textbf{P1.\ Sensing \& Data Intelligence}\\
  \textit{Observe phase}\\[3pt]
  \scriptsize
  Multi-modal agents (SCADA, IoT, vision, EWS) $\cdot$
  novelty / zero-day detection $\cdot$ cyber--physical correlation
  $\cdot$ federated learning $\cdot$ blockchain-secured CTI
  (STIX/TAXII) $\cdot$ environmental EWS
  \end{minipage}
};
\node[pillar,fill=rrep!18] (p2) at (0,-2.55) {
  \begin{minipage}{6.0cm}\centering
  \textbf{P2.\ Resilience Analytics \& Reporting}\\
  \textit{Simulate \& Decide phase}\\[3pt]
  \scriptsize
  DTaaS (INTO-CPS) $\cdot$ cyber--physical co-simulation $\cdot$
  multi-hazard what-if contingency $\cdot$ AI risk assessment \&
  cascading impact $\cdot$ resilience metrics \& dashboards
  \end{minipage}
};
\node[pillar,fill=rsvc!18] (p3) at (0,-5.1) {
  \begin{minipage}{6.0cm}\centering
  \textbf{P3.\ Incident Response \& Recovery}\\
  \textit{Act \& Adapt phase}\\[3pt]
  \scriptsize
  SOAR playbooks with LLM forensic rationale $\cdot$ adaptive
  containment $\cdot$ self-healing, islanding, failover $\cdot$
  service migration \& business continuity $\cdot$ blockchain audit
  $\cdot$ CER/NIS2 reporting
  \end{minipage}
};
\node[hub] (hub) at (0,-7.35) {RESILAGENT Cloud-based Orchestrator\\
  SOC $\cdot$ NOC $\cdot$ Energy Control Room\\
  \footnotesize(mediation, ranking, SOAR, escalation, blockchain evidence vault)};

\draw[arr] (p1.south) -- (p2.north);
\draw[arr] (p2.south) -- (p3.north);
\draw[arr] (p3.south) -- (hub.north);
\coordinate (rail) at ($(p1.west)+(-0.7,0)$);
\draw[thick,black!65] (rail) -- (rail |- hub.west) -- (hub.west);
\foreach \p in {p1,p2,p3}{
  \draw[arr] (rail |- \p.west) -- (\p.west);
}
\node[rotate=90,font=\scriptsize\itshape,fill=white,inner sep=1pt]
  at ($(rail)!0.5!(rail |- hub.west)$) {Orchestrator mediation};
\draw[arr,dashed,black!55] (p3.east) to[out=0,in=0,looseness=1.6]
  node[right,font=\scriptsize\itshape,align=left]
  {continuous\\resilience\\loop} (p1.east);

\begin{scope}[on background layer]
\node[outer,fit={(p1) (p2) (p3) (hub) (rail)}] (frame) {};
\end{scope}
\node[anchor=south,font=\bfseries\small,
      fill=white,inner xsep=8pt,inner ysep=1pt]
  at (frame.north) {RESILAGENT framework};
\end{tikzpicture}%
}
\caption{RESILAGENT three-pillar architecture: P1 Sensing, P2
Analytics, P3 Response, mediated by a Cloud-based Orchestrator.
Solid arrows trace one workflow cycle; the dashed arrow closes the
loop.}
\label{fig:arch}
\end{figure}
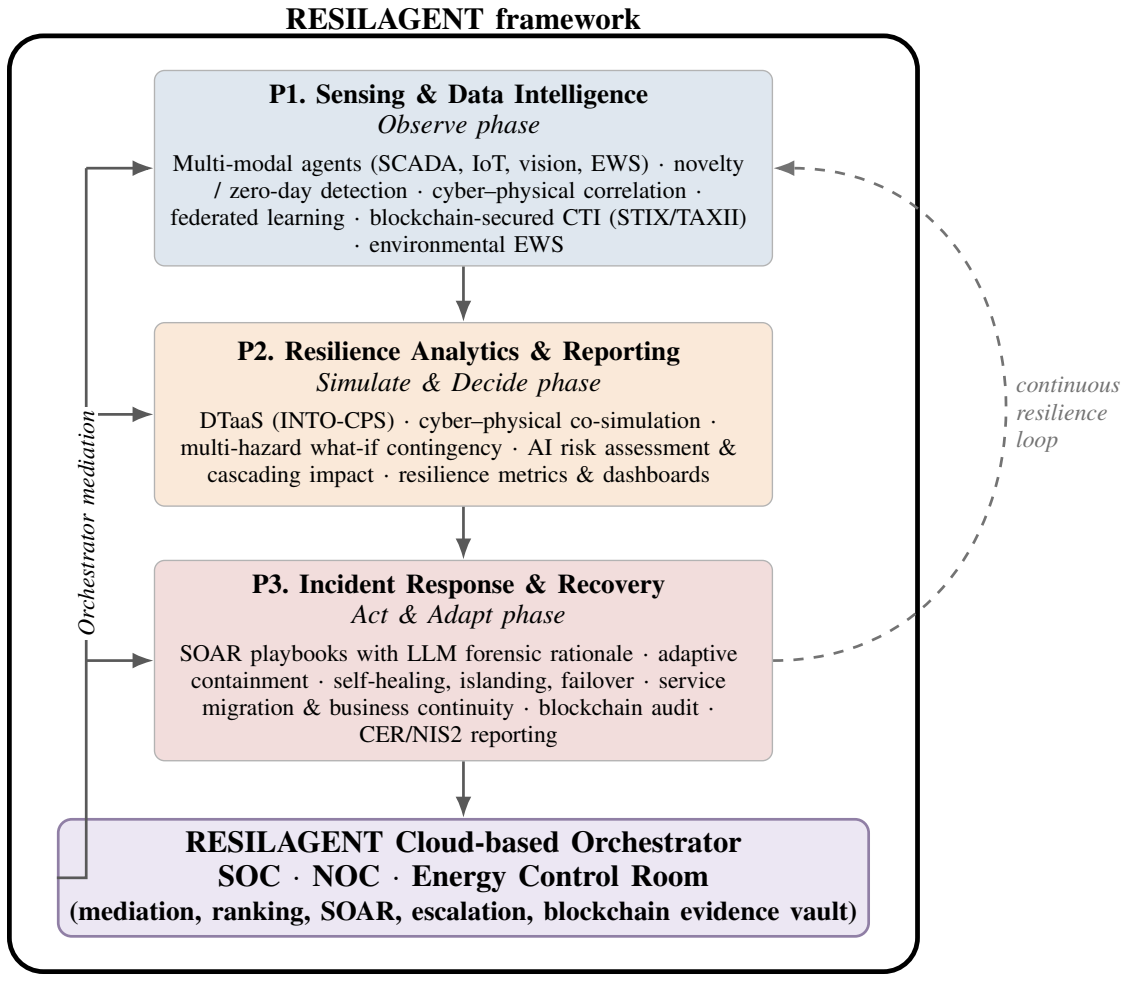

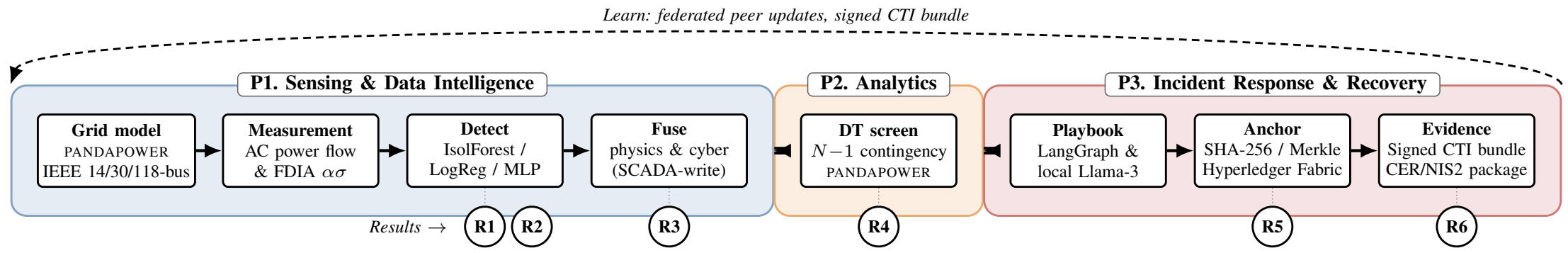
\begin{figure*}[t]
\centering
\resizebox{0.80\textwidth}{!}{%
\begin{tikzpicture}[
  font=\scriptsize,
  stage/.style={draw=black,thick,rectangle,rounded corners=2pt,
                minimum width=2.15cm,minimum height=1.0cm,
                align=center,inner sep=2pt,fill=white},
  group/.style={draw=black!65,thick,rounded corners=6pt,
                inner xsep=10pt,inner ysep=11pt,fill=black!5},
  grouplbl/.style={font=\bfseries\footnotesize,
                   fill=white,inner xsep=5pt,inner ysep=1pt,
                   draw=black!65,thin,rounded corners=2pt},
  resultlbl/.style={draw=black,thick,circle,
                    minimum size=0.55cm,align=center,
                    font=\scriptsize\bfseries,inner sep=0pt,
                    fill=white},
  flow/.style={-{Latex[length=2.4mm,width=2.0mm]},very thick,black},
  flowCross/.style={-{Latex[length=2.4mm,width=2.0mm]},
                    very thick,black,double},
  back/.style={-{Latex[length=2.4mm,width=2.0mm]},thick,black,
               densely dashed},
  rline/.style={black!50,thin,densely dotted},
]

\node[stage] (s1) at (0,0)
  {\textbf{Grid model}\\\textsc{pandapower}\\IEEE 14/30/118-bus};
\node[stage] (s2) at (2.55,0)
  {\textbf{Measurement}\\AC power flow\\\& FDIA $\alpha\sigma$};
\node[stage] (s3) at (5.10,0)
  {\textbf{Detect}\\IsolForest /\\LogReg / MLP};
\node[stage] (s4) at (7.65,0)
  {\textbf{Fuse}\\physics \& cyber\\(SCADA-write)};

\node[stage] (s5) at (10.55,0)
  {\textbf{DT screen}\\$N{-}1$ contingency\\\textsc{pandapower}};

\node[stage] (s6) at (13.45,0)
  {\textbf{Playbook}\\LangGraph \&\\local Llama-3};
\node[stage] (s7) at (16.00,0)
  {\textbf{Anchor}\\SHA-256 / Merkle\\Hyperledger Fabric};
\node[stage] (s8) at (18.55,0)
  {\textbf{Evidence}\\Signed CTI bundle\\CER/NIS2 package};

\begin{scope}[on background layer]
  \node[group,fill=ring!14,draw=ring!70,fit={(s1) (s2) (s3) (s4)}] (P1box) {};
  \node[group,fill=rrep!14,draw=rrep!70,fit=(s5)] (P2box) {};
  \node[group,fill=rsvc!14,draw=rsvc!70,fit={(s6) (s7) (s8)}] (P3box) {};
\end{scope}

\node[grouplbl] at (P1box.north)
  {P1.~Sensing \& Data Intelligence};
\node[grouplbl] at (P2box.north) {P2.~Analytics};
\node[grouplbl] at (P3box.north)
  {P3.~Incident Response \& Recovery};

\foreach \i/\j in {s1/s2,s2/s3,s3/s4,s6/s7,s7/s8}{
  \draw[flow] (\i.east) -- (\j.west);
}
\draw[flowCross] (P1box.east) -- (P2box.west);
\draw[flowCross] (P2box.east) -- (P3box.west);

\draw[back] (P3box.north east) to[out=110,in=70,looseness=0.12]
  node[above,font=\scriptsize\itshape,pos=0.5]
  {Learn: federated peer updates, signed CTI bundle}
  (P1box.north west);

\node[resultlbl] (r1) at (s3.south |- 0,-1.05) {R1};
\node[resultlbl] (r2) at ($(s3.south |- 0,-1.05)+(0.65,0)$) {R2};
\node[resultlbl] (r3) at (s4.south |- 0,-1.05) {R3};
\node[resultlbl] (r4) at (s5.south |- 0,-1.05) {R4};
\node[resultlbl] (r5) at (s7.south |- 0,-1.05) {R5};
\node[resultlbl] (r6) at (s8.south |- 0,-1.05) {R6};
\draw[rline] (s3.south) -- (r1.north);
\draw[rline] (s4.south) -- (r3.north);
\draw[rline] (s5.south) -- (r4.north);
\draw[rline] (s7.south) -- (r5.north);
\draw[rline] (s8.south) -- (r6.north);
\node[font=\scriptsize\itshape,anchor=east]
  at (r1.west) [xshift=-3pt] {Results $\to$};

\end{tikzpicture}%
}
\caption{End-to-end experimental pipeline. Eight stages grouped into
three pillar containers: P1 prepares the grid model and runs
detection plus cyber--physical fusion; P2 screens $N{-}1$
contingencies on the DT; P3 generates the LLM SOAR playbook, anchors
artefacts and emits the CER/NIS2 evidence package. Badges R1--R6
mark the results in Section~\ref{sec:eval} (R6 ties the chain
end-to-end).}
\label{fig:pipeline}
\end{figure*}

Figure~\ref{fig:arch} shows the three pillars mediated by
the Cloud-based Orchestrator;
Figure~\ref{fig:pipeline} deepens this into the experimental pipeline
exercised by the micro-benchmark
(Section~\ref{sec:eval}) and the two pilots (Section~\ref{sec:scenarios}).

\subsection{Minimum Viable Stack}\label{ssec:mvs}
Only four technologies are \emph{irreducible} for a first deployment:
\emph{(i)}~a multi-modal monitoring \textbf{agent} with at least one
cyber and one physical input (P1); \emph{(ii)}~a real-time
\textbf{Digital Twin} supplying the residual / contingency reference
frame (P2); \emph{(iii)}~an \textbf{LLM-backed fusion / playbook agent}
binding heterogeneous inputs into explainable actions (P1$\to$P3); and
\emph{(iv)}~a \textbf{blockchain anchor} signing every detection,
simulation and action (P3). Removing any one breaks the loop. Every
other technology in Fig.~\ref{fig:arch}, vision-LLM, federated
learning, EWS, blockchain-anchored CTI, service/data migration,
compliance automation, AI risk scoring, is a stageable
\emph{extension}; we recommend the four-core stack on a single-sector
pilot first, then extensions as operator maturity and regulatory
pressure require.

\subsection{The Three Pillars and the Orchestrator}
\textit{P1, Sensing \& Data Intelligence.} Modality-specific agents
observe SCADA, Phasor Measurement Unit (PMU), video, acoustic,
vibration and environmental Early-Warning-System (EWS)
sensors~\cite{wu2025graphkan,saber2026}; novelty detection targets
zero-day patterns; a multimodal LLM fusion agent binds anomalies
sharing site, time or asset into hybrid-event
hypotheses~\cite{zaboli2024}; federated learning improves detection
across operators without sharing raw data~\cite{idsa_energy2024}; and
a blockchain-secured CTI plane enables tamper-evident peer exchange
under the STIX/TAXII standards~\cite{tounsi2018,mdpi_bccti2024,spear2021}.
\textit{P2, Resilience Analytics \& Reporting.} A DTaaS layer
(INTO-CPS)~\cite{intocps2016} hosts high-fidelity grid twins for predictive simulation
and stress testing~\cite{eckhart2019}; co-simulation supports
real-time cyber-physical reconfiguration~\cite{saad2020,helics2017,temperekidis2022};
multi-hazard simulation combines cyber, physical and environmental
stressors; AI-driven risk assessment evaluates cascading impacts; and
dashboards consolidate recommendations, on an
open stack of \textsc{pandapower}~\cite{pandapower2018},
HELICS and OPAL-RT (HIL).
\textit{P3, Incident Response \& Recovery.} Once P2 ranks mitigations,
P3 runs AI-assisted forensics and adaptive containment, LLM-based
explainable playbooks~\cite{sharshar2025}, survivability via
hot standby, automated failover and intentional islanding,
immutable blockchain audit~\cite{tounsi2018}, and CER/NIS2
compliance automation~\cite{cer2022,nis2}.
\textit{Orchestrator.} A Cloud-based Orchestrator presents a unified
SOC / Network Operations Centre (NOC) / Energy Control Room view,
ranks P2 mitigations, automates low-risk SOAR responses, escalates to
civil-protection authorities, and signs every action and CTI artefact
in a blockchain-anchored evidence vault~\cite{nis2,cer2022}.

\subsection{Closed-Loop Resilience Workflow}
\label{ssec:workflow}
At runtime the pillars interact through a single closed loop,
the eight stages drawn in
Fig.~\ref{fig:pipeline} and closed by the dashed return arrow of
Fig.~\ref{fig:arch}: \emph{Observe} (P1) detects and correlates per-modality
anomalies;
\emph{Simulate \& Decide} (P2) explores contingency paths on the live
DT and returns a ranked, confidence-scored mitigation list;
\emph{Act \& Adapt} (P3) executes low-risk actions through SOAR while
high-risk actions await operator confirmation, then
\emph{learns}, emitting a blockchain-anchored CER/NIS2 evidence package
that feeds both the next cycle and external auditors.

\subsection{Deployment View}
\label{ssec:deploy}
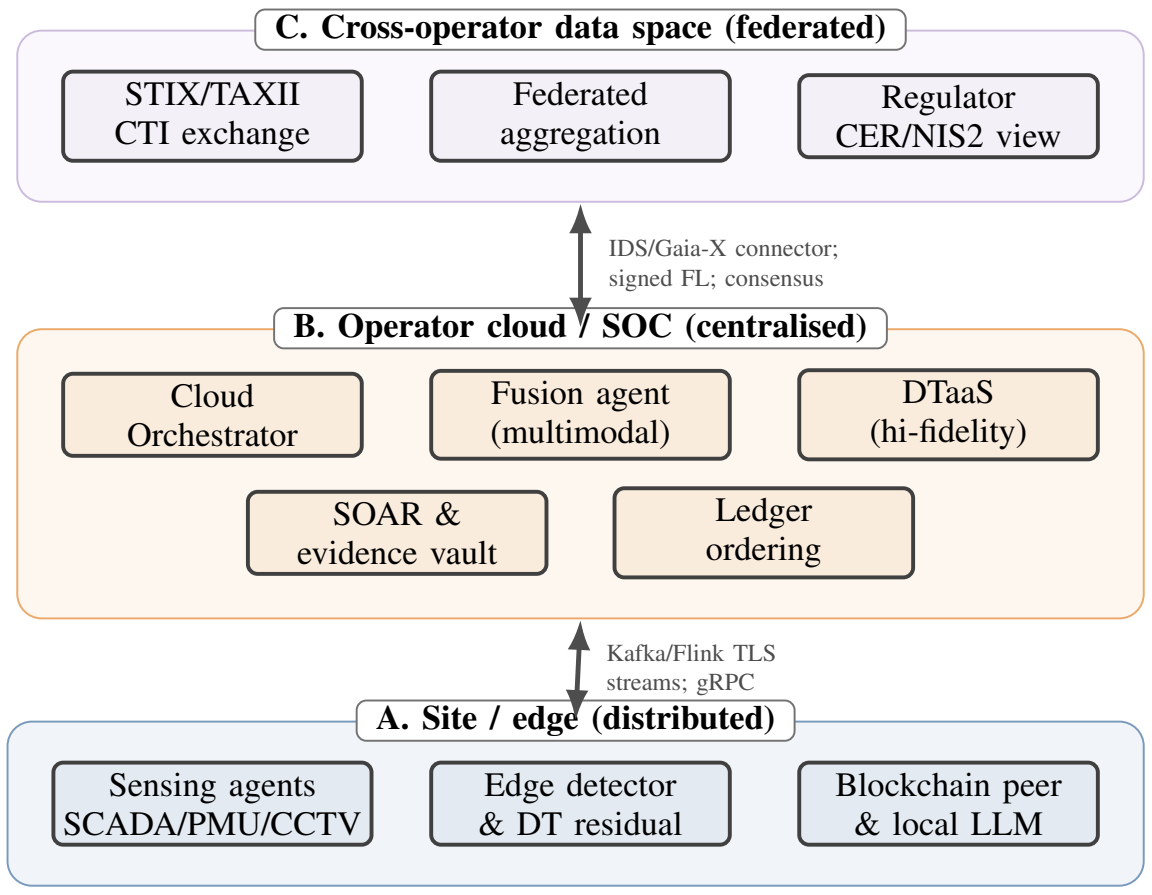
\begin{figure}[t]
\centering
\resizebox{0.85\columnwidth}{!}{%
\begin{tikzpicture}[
  font=\scriptsize,
  comp/.style={draw=black!75,thick,rounded corners=2pt,fill=white,
               minimum height=0.55cm,minimum width=2.15cm,
               inner sep=2pt,align=center},
  band/.style={rounded corners=5pt,inner xsep=9pt,inner ysep=8pt},
  tlbl/.style={font=\bfseries\scriptsize,fill=white,inner xsep=4pt,
               inner ysep=1pt,draw=black!55,thin,rounded corners=2pt},
  link/.style={{Latex[length=2.2mm]}-{Latex[length=2.2mm]},
               line width=1pt,black!70},
]
\node[comp,fill=rblk!16] (cti) at (-2.65,4.15) {STIX/TAXII\\CTI exchange};
\node[comp,fill=rblk!16] (fl)  at ( 0.00,4.15) {Federated\\aggregation};
\node[comp,fill=rblk!16] (reg) at ( 2.65,4.15) {Regulator\\CER/NIS2 view};
\begin{scope}[on background layer]
  \node[band,draw=rblk!70,fill=rblk!7,fit=(cti)(fl)(reg)] (Cband) {};
\end{scope}
\node[tlbl] at (Cband.north) {C. Cross-operator data space (federated)};

\node[comp,fill=rrep!16] (orch)  at (-2.65,2.0) {Cloud\\Orchestrator};
\node[comp,fill=rrep!16] (fuse)  at ( 0.00,2.0) {Fusion agent\\(multimodal)};
\node[comp,fill=rrep!16] (dtaas) at ( 2.65,2.0) {DTaaS\\(hi-fidelity)};
\node[comp,fill=rrep!16] (soar)  at (-1.32,1.15) {SOAR \&\\evidence vault};
\node[comp,fill=rrep!16] (order) at ( 1.32,1.15) {Ledger\\ordering};
\begin{scope}[on background layer]
  \node[band,draw=rrep!70,fill=rrep!7,
        fit=(orch)(fuse)(dtaas)(soar)(order)] (Bband) {};
\end{scope}
\node[tlbl] at (Bband.north) {B. Operator cloud / SOC (centralised)};

\node[comp,fill=ring!16] (sens) at (-2.65,-0.8) {Sensing agents\\SCADA/PMU/CCTV};
\node[comp,fill=ring!16] (edge) at ( 0.00,-0.8) {Edge detector\\\& DT residual};
\node[comp,fill=ring!16] (peer) at ( 2.65,-0.8) {Blockchain peer\\\& local LLM};
\begin{scope}[on background layer]
  \node[band,draw=ring!75,fill=ring!8,fit=(sens)(edge)(peer)] (Aband) {};
\end{scope}
\node[tlbl] at (Aband.north) {A. Site / edge (distributed)};

\draw[link] (Aband.north) -- node[right,font=\tiny,align=left,xshift=2pt]
  {Kafka/Flink TLS\\streams; gRPC} (Bband.south);
\draw[link] (Bband.north) -- node[right,font=\tiny,align=left,xshift=2pt]
  {IDS/Gaia-X connector;\\signed FL; consensus} (Cband.south);
\end{tikzpicture}%
}
\caption{RESILAGENT deployment view. Tier~A (edge, distributed) keeps
raw telemetry on-premise; Tier~B (operator cloud / SOC, centralised)
hosts orchestration, fusion, DTaaS, SOAR and the single
human-in-the-loop checkpoint; Tier~C (cross-operator data space,
federated) shares only model updates and signed CTI. Arrows mark the
inter-tier communication paths and protocols.}
\label{fig:deploy}
\end{figure}

Fig.~\ref{fig:deploy} maps the framework onto three tiers: a
\emph{distributed} edge that keeps raw SCADA on-premise (sensing
agents, a DT-residual detector, a blockchain peer with optional local
LLM); a \emph{centralised} operator cloud / SOC hosting the
Orchestrator, fusion agent, DTaaS, SOAR and the single
human-in-the-loop checkpoint; and a \emph{federated} tier sharing only
model updates and signed CTI over the permissioned ledger.
Edge-to-cloud traffic uses TLS-secured Kafka/Flink and gRPC;
cloud-to-data-space an International-Data-Spaces / Gaia-X connector.

\section{EPES Application Scenarios}\label{sec:scenarios}

We ground the framework in two pilot scenarios, a national TSO and a
regional DSO, each illustrating one stress mode; every pilot step maps
to a result of Section~\ref{sec:eval}.

\subsection{TSO Scenario: Cyber Intrusion under a Heatwave}
\label{ssec:tso}
A national TSO with ${>}9\,000$~km of high-voltage lines and
ENTSO-E interconnections faces a coordinated hybrid campaign during a
42\,$^\circ$C summer peak that combines (i)~credential compromise on a
SCADA VPN gateway, (ii)~deliberate damage at a 400~kV transformer
substation, and (iii)~a heatwave reducing hydro output by 40\,\%.
Under RESILAGENT, P1 fuses an Operational-Technology (OT) flow anomaly
(R1--R3-class), a vision-LLM signal on substation CCTV and a
Sentinel-2 tile, and an EWS heatwave alert into one
hypothesis; P2 runs the $N{-}1$ contingency on a Common-Information-Model
network in \textsc{pandapower} (R4-class), finding that
severing one 400~kV line overloads two parallel ones; P3 emits an
LLM SOAR playbook (R6-class) proposing load shedding,
genset dispatch and VPN-segment isolation, all signed by the
anchor (R5).

\subsection{DSO Scenario: DER Supply-Chain Compromise under a Storm}
\label{ssec:dso}
A regional DSO with growing DER penetration is exposed to a
sector-wide supply-chain risk on DER firmware~\cite{enisa2024,enisa2025}.
A malicious update is silently installed on a cluster of solar
inverters months before an Atlantic storm; at landfall, wind brings
down feeders and the implant triggers, disconnecting tens of megawatts
of inverter capacity. Under RESILAGENT, the EWS agent pre-stages the
DT with the storm trajectory; a
firmware-attestation agent hashes fielded firmware against the
blockchain-anchored vendor-advisory ledger~\cite{tounsi2018} and flags
the malicious version before it fires; at impact, the fusion agent
combines storm position, mass inverter dropout (SCADA/DER telemetry)
and post-event imagery from a vision-LLM damage
classifier; the DT re-runs with the updated topology (R4-class) and
the P3 agent (R6-class) islands the segment serving a hospital and
shelter, every step bound to the same SHA-256 chain (R5)
producing the CER/NIS2 evidence package~\cite{cer2022,nis2}.

\section{Evaluation Roadmap}\label{sec:eval}

\subsection{Open-Source Toolchain and Datasets}
RESILAGENT is being prototyped on an open-source stack:
\textsc{pandapower}~\cite{pandapower2018} for the static
power-flow DT; HELICS~\cite{helics2017} for cyber--physical
co-simulation; OPAL-RT for HIL where licensed; Apache Kafka/Flink for
the ingestion plane; Suricata, Zeek and Wazuh for cyber sensing;
LangGraph for the agent layer with locally hosted LLMs
(Llama-3, Mistral) for sovereignty; and Hyperledger
Fabric for the trust anchor. Planned benchmark datasets include PNNL
Electricity/Gas IDS, HAI~22.04, iTrust EPIC, ENTSO-E Transparency,
OEDI outage events and Copernicus EWS feeds.

\subsection{Metrics}
Beyond detection statistics (F1, precision, recall, per
modality and fused), the project will report MTTD and MTTR,
\emph{service-continuity} (kW and critical-load fraction
restored), \emph{operator burden} (analyst-minutes per incident), a
composite \emph{resilience index} over severity, duration and recovery,
and \emph{compliance latency} from incident
onset to a NIS2-compliant evidence package~\cite{nis2}.

\begin{figure}[t]
\centering
\begin{minipage}[t]{0.49\columnwidth}
  \centering
  \includegraphics[width=0.88\linewidth]{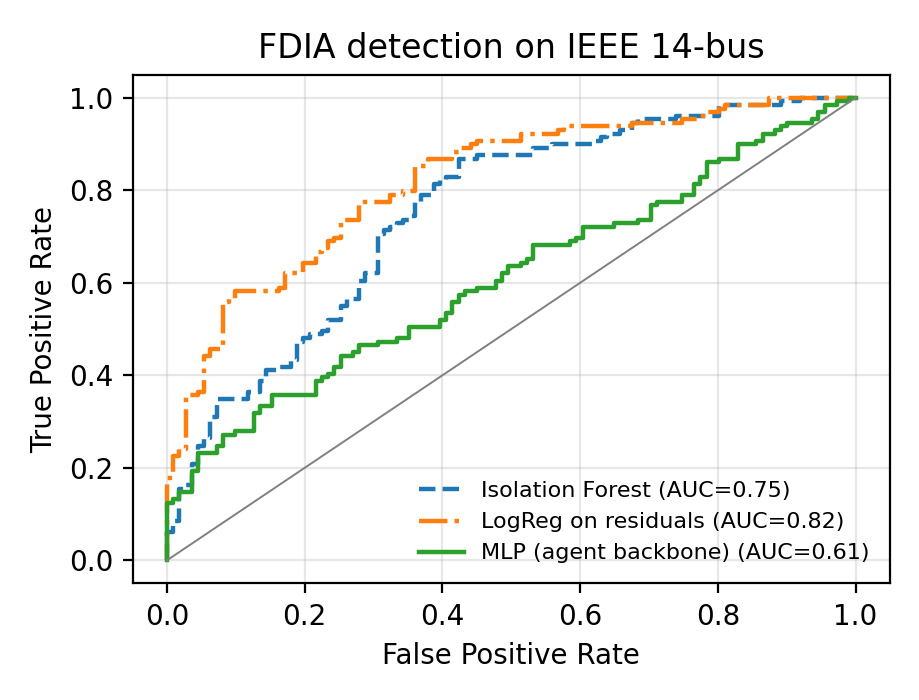}\\
  {\scriptsize (a) ROC, $\alpha{=}2\sigma$}
\end{minipage}\hfill
\begin{minipage}[t]{0.49\columnwidth}
  \centering
  \includegraphics[width=0.88\linewidth]{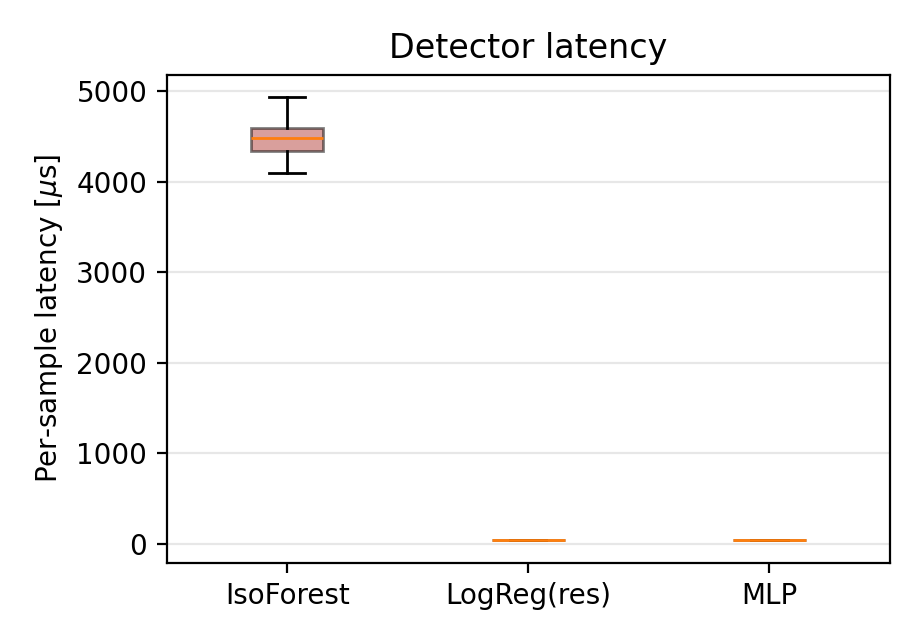}\\
  {\scriptsize (b) Inference latency}
\end{minipage}

\vspace{3pt}
\begin{minipage}[t]{0.49\columnwidth}
  \centering
  \includegraphics[width=0.88\linewidth]{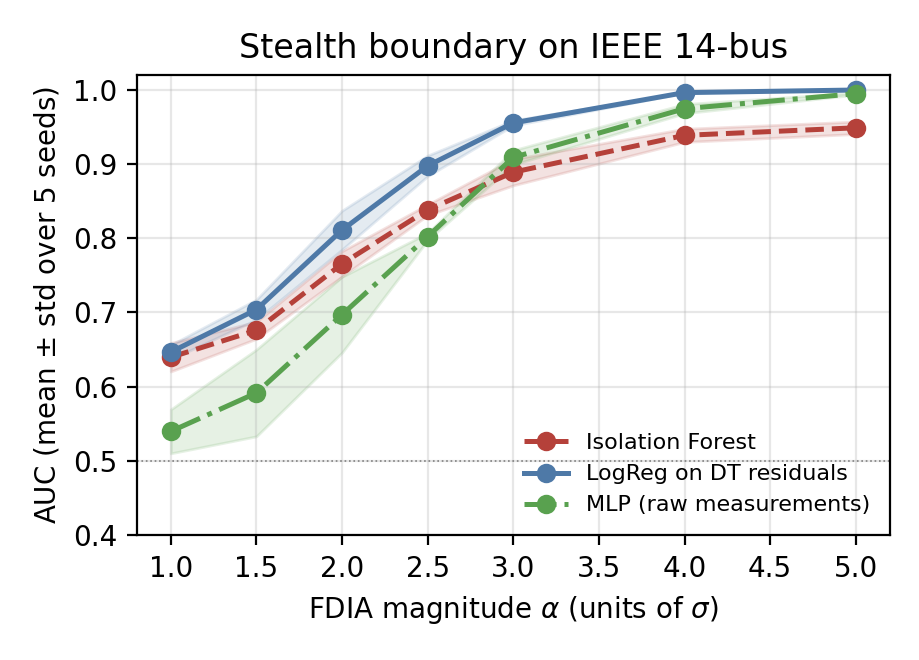}\\
  {\scriptsize (c) AUC vs stealth $\alpha$}
\end{minipage}\hfill
\begin{minipage}[t]{0.49\columnwidth}
  \centering
  \includegraphics[width=0.88\linewidth]{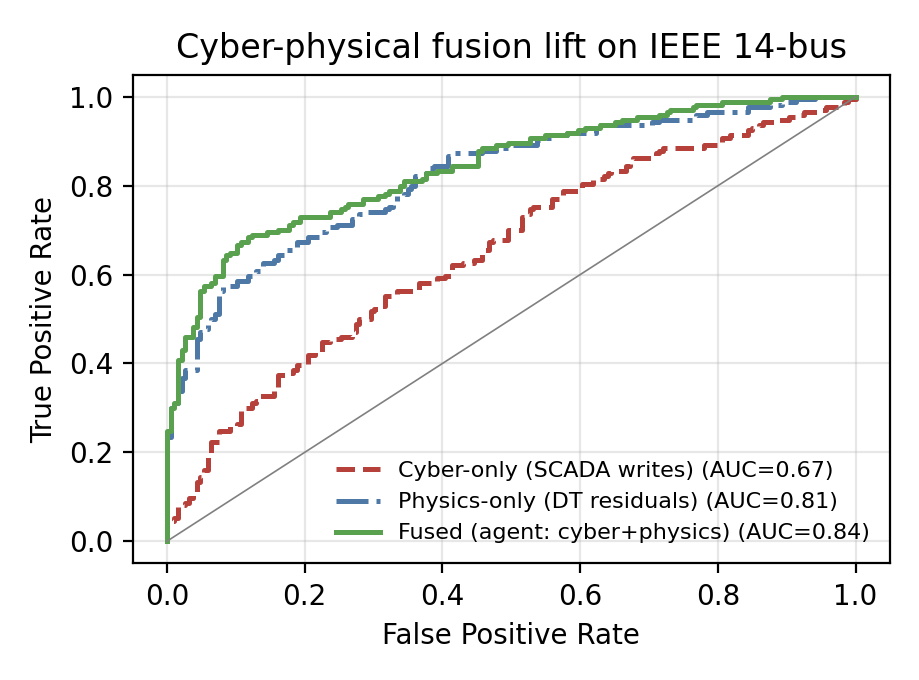}\\
  {\scriptsize (d) Cyber--physical fusion}
\end{minipage}
\caption{P1 micro-benchmark on IEEE 14-bus:
(a)~ROC at $\alpha{=}2\sigma$ (2 buses, 800 snapshots);
(b)~per-sample inference latency;
(c)~stealth-boundary sweep, mean$\pm$std (3 seeds);
(d)~cyber--physical fusion AUC (5 seeds).}
\label{fig:p1}
\end{figure}

\subsection{Initial Micro-Benchmark}
\label{ssec:results}

\paragraph*{Setup}
All experiments run on a 2024 Apple-silicon laptop (Python~3.9,
\textsc{pandapower}~3.2.2), complete in under five minutes
on deterministic seeds, and total roughly 500 lines of
Python. The IEEE 14-, 30- and 118-bus
cases~\cite{pandapower2018} provide the grid models;
for each snapshot we jitter every load by $\pm 25\,\%$, solve AC power
flow to obtain the true bus-voltage vector $\mathbf{x}$, and
synthesise the measurement vector
$\mathbf{z}=\mathbf{x}+\mathcal{N}(0,\sigma^2)$, where
$\mathcal{N}(0,\sigma^2)$ denotes zero-mean Gaussian noise with
standard deviation $\sigma{=}0.01$ in per-unit (p.u.). FDIA samples
add an offset drawn from the continuous uniform distribution
$\mathcal{U}[\alpha{-}1,\alpha{+}1]\,\sigma$ on $k$ randomly chosen
buses, where the stealth parameter $\alpha$ scales the injection (a
smaller $\alpha$ is stealthier) and $k$ is the number of attacked
buses. We contrast three detectors that stand in for three
architectural choices: an unsupervised \emph{Isolation Forest} (no
labels, no physics); a \emph{Logistic Regression on residual
statistics} (the $\max$, mean and standard deviation of the residuals
$r_i$, and the count of $|r_i|{>}3$, where $r_i$ is the per-bus
measurement residual) modelling the physics-aware signal a P1 agent
extracts via the DT; and a small \emph{Multi-Layer Perceptron (MLP)}
on the raw measurement vector, modelling a feature-blind end-to-end
neural detector.

\paragraph*{Scope}
The benchmark is intentionally a \emph{proxy evaluation}: it does not
claim to validate a national TSO or DSO deployment. Each experiment
isolates one mechanism required by the proposed architecture
(physics-aware detection, cyber--physical feature fusion, DT
contingency screening, cryptographic anchoring, and orchestration
latency); the cyber feature, the vision input and the production
ledger are deliberately replaced by lightweight proxies. Detection and
fusion (Results~1--3) are reported on the IEEE 14-bus case, while DT
contingency screening (Result~4) and anchor cost (Result~5) extend
directly to the 30- and 118-bus networks; a full multi-case
characterisation of detection and fusion across all three networks is
the immediate next step.

\begin{figure}[t]
\centering
\begin{minipage}[t]{0.49\columnwidth}
  \centering
  \includegraphics[width=\linewidth]{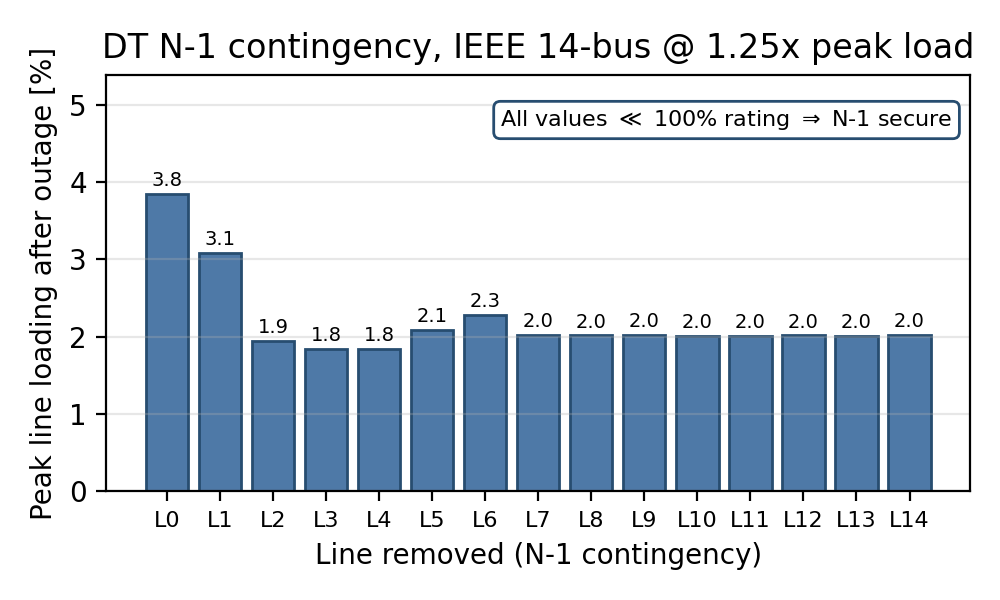}\\
  {\scriptsize (a) P2: $N{-}1$ DT screen}
\end{minipage}\hfill
\begin{minipage}[t]{0.49\columnwidth}
  \centering
  \includegraphics[width=\linewidth]{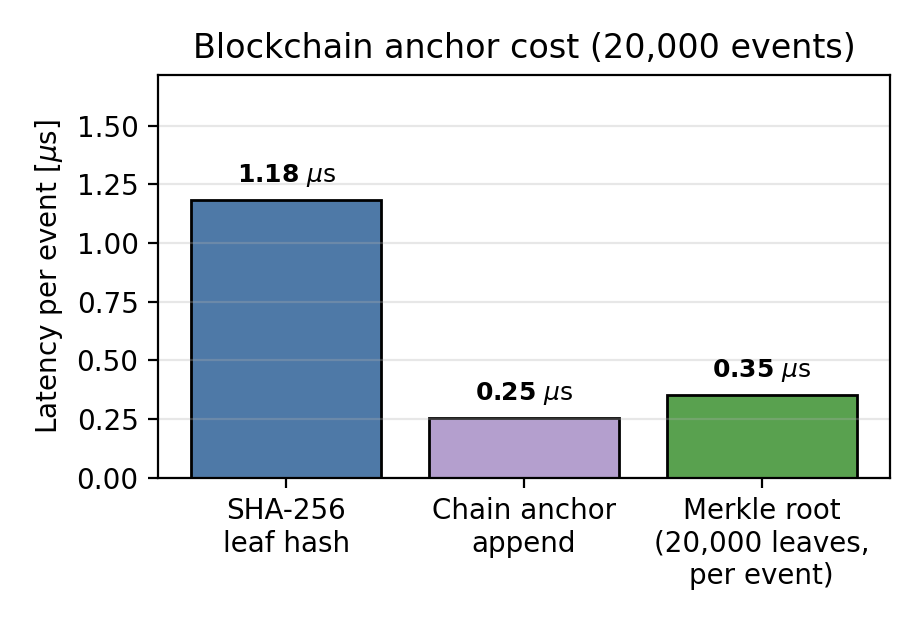}\\
  {\scriptsize (b) P3: anchor cost}
\end{minipage}

\vspace{3pt}
\includegraphics[width=0.60\columnwidth]{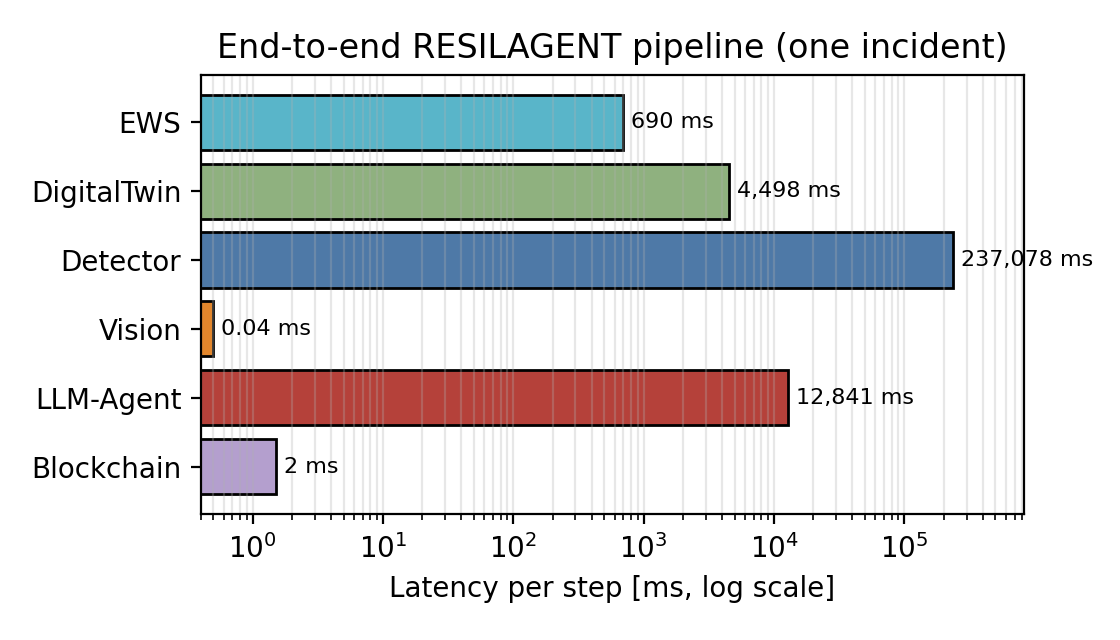}\\
{\scriptsize (c) Result~6: end-to-end pipeline, one incident}
\caption{Quantitative results. (a)~P2 $N{-}1$ contingency screen on
IEEE 14-bus at $1.25\times$ peak load (15 contingencies, audit
$\sim 250$\,ms); (b)~P3 anchor cost over 20\,000 events,
sub-microsecond per event and three orders of magnitude below detector
latency; (c)~Result~6 end-to-end pipeline, six tools in sequence on
commodity hardware. The \textsc{Detector} bar is inflated because the
benchmark retrains from scratch; in deployment that step is
sub-millisecond.}
\label{fig:p2p3}
\vspace{-10pt}
\end{figure}

\paragraph*{Pillar 1, Result~1: nominal operating point}
At $\alpha{=}2\sigma$ on two random buses (800 snapshots), the
residual-aware Logistic Regression reaches an Area Under the
Receiver-Operating-Characteristic Curve (AUC)~$=0.82$, the
unsupervised Isolation Forest AUC~$=0.75$, and the feature-blind MLP
collapses to AUC~$=0.61$ (Fig.~\ref{fig:p1}(a)). Median per-sample
latency (Fig.~\ref{fig:p1}(b)) is 48\,$\mu$s (LogReg), 43\,$\mu$s
(MLP) and 4.5\,ms (Isolation Forest), all within sub-second
control-loop budgets. The DT-derived $z$-scored residuals carry the
inductive bias a feature-blind network cannot learn at this sample size.

\paragraph*{Pillar 1, Result~2: stealth boundary}
Sweeping $\alpha\in[1,5]\sigma$ over three seeds
(Fig.~\ref{fig:p1}(c)), the residual-aware detector dominates
everywhere and reaches AUC~$\geq 0.95$ from $\alpha{\geq}3\sigma$,
while the MLP fails in the stealth regime (AUC~$0.54$ at
$\alpha{=}1\sigma$) and only catches up at $\alpha{\geq}4\sigma$ where
a human is likely to notice anyway. The gap is widest where the
signal is weakest, the practically relevant regime for advanced
adversaries.

\paragraph*{Pillar 1, Result~3: cyber--physical fusion}
With a synthetic 1-D ``SCADA write-rate'' feature (Gaussian-shifted
by ${\sim}0.7\sigma$ under attack) and Logistic Regression over five
seeds (1\,200 snapshots), AUCs are physics-only $0.807\pm 0.011$,
cyber-only $0.675\pm 0.033$,
\textbf{fused $\mathbf{0.838\pm 0.003}$} (Fig.~\ref{fig:p1}(d)).
Fusion lifts AUC by 3.1\,pp over physics-only and 16.3\,pp over
cyber-only, and the seed-to-seed standard deviation drops by an
order of magnitude. That stability is the empirical case for the
multimodal fusion agent at the heart of P1.

\paragraph*{Pillar 2, Result~4: Digital-Twin contingency screening}
On IEEE 14-bus at $1.25\times$ peak load the DT enumerates all 15
$N{-}1$ contingencies in $\sim 250$\,ms; none exceeds the 100\,\%
rating (Fig.~\ref{fig:p2p3}(a)). The loadings are mild on this
generously rated case; the point is that a complete sub-second
$N{-}1$ audit turns an annual offline study into a \emph{live}
dashboard, the quantitative ground that lets P3 propose pre-emptive
load shedding rather than react to a trip.

\paragraph*{Pillar 3, Result~5: blockchain anchor cost}
We stream 20\,000 synthetic incident events (each carrying a 32-byte
payload hash, operator id, type and timestamp) through three signing
schemes implemented with Python \texttt{hashlib}: a per-event
SHA-256 leaf, an append to a linear chain, and a Merkle tree over
the whole window. Averaged cost per event is
1.20\,$\mu$s (SHA-256 leaf), 0.26\,$\mu$s (chain
append) and 0.35\,$\mu$s (Merkle root)
(Fig.~\ref{fig:p2p3}(b)), three orders of magnitude below the detector
latency (${\sim}50\,\mu$s) and four below the DT solve
(${\sim}16$\,ms): local anchoring is negligible at the control-loop
budget. A production permissioned ledger (Hyperledger Fabric) would
add millisecond-range ordering, endorsement and commit latency, which
remains deployment-dependent.

\paragraph*{Result~6: end-to-end pipeline}
To show the components compose, we chain \emph{all} tools in a
single-incident script: a live Open-Meteo EWS fetch, the IEEE~14-bus
DT $N{-}1$ audit triggered by the EWS regime label, the FDIA detector
of Results~1--3, a vision stub, a local Llama-3.2 agent emitting a
JSON action plan, and a SHA-256 chain anchoring every step. The agent
labelled the regime \textsc{STORM}, the DT cleared the network as
$N{-}1$-secure, and the agent emitted an ordered high-severity plan
(12.8\,s), anchored to a deterministic root. We deliberately separate
\emph{proxy artefacts} from \emph{deployment estimates}: the 255\,s
wall-clock is dominated by a benchmark detector that retrains from
scratch (the 237\,s \textsc{Detector} bar, Fig.~\ref{fig:p2p3}(c)),
absent in deployment where inference is the 50\,$\mu$s of Result~1.
The deployment-representative path, the DT $N{-}1$ audit
(${\sim}250$\,ms), the Llama-3.2 rationale (12.8\,s) and
microsecond-scale detection and anchoring, gives an effective
LLM~\&~DT decision latency of ${\sim}17$\,s, wrong for
protection-relay loops but right for the human-in-the-loop stage the
framework targets.

\subsection{Federated Learning and Continuous Risk}\label{ssec:fl}
\paragraph*{Federated learning across operators}
Federated learning (FL) is the mechanism by which detection improves
across the European fleet without raw data ever leaving an operator,
directly serving the data-residency obligations of NIS2 and the EU
data-space lineage~\cite{idsa_energy2024} (e.g.\
\href{https://cordis.europa.eu/project/id/101069831}{ENERSHARE}). Each
TSO/DSO trains the multimodal anomaly model locally and shares only
model updates; a secure-aggregation round (FedAvg-style) at the
centralised aggregator of Tier~B (Fig.~\ref{fig:deploy}) combines them
into a new global model. Because the threat landscape is shared but
telemetry is not, FL lets a stealth-FDIA pattern seen by one operator
harden every peer's detector while each keeps its SCADA on-premise.
The open risk is model poisoning by a compromised participant or
sensor; RESILAGENT mitigates it by signing every update and anchoring
the round on the trust ledger (Result~5), so a poisoned contribution
is attributable and revocable, and by DT-based plausibility filtering
of the aggregated model. A quantitative FL evaluation, convergence and
poisoning robustness across operators, is a primary item on the pilot
roadmap.

\paragraph*{Continuous risk and resilient operations}
RESILAGENT replaces the annual-audit posture with a per-event loop
that re-scores threat likelihood (P1), impact (DT $N{-}1$/$N{-}2$,
Result~4) and residual mitigation risk continuously. A Governance,
Risk and Compliance cockpit maps every artefact to CER and NIS2
controls~\cite{nis2,cer2022}, and the anchor (Result~5) makes the
chain externally verifiable, dropping prototype evidence generation
to seconds.
Supply-chain and AI-model risk, the blind spot of cyber-only
frameworks, is covered by firmware attestation against the anchored
vendor-advisory ledger (DSO pilot) and the signed FL updates above.

\section{Open Challenges}\label{sec:challenges}

\textit{Trust, explainability and real-time performance.} Operators
will not accept agent escalations they cannot interrogate, and EPES
control loops are sub-second while large LLM inference is not.
RESILAGENT pairs every recommendation with a DT-derived rationale
(Result~4) and a signed evidence trail, and tiers compute: distilled
models on the hot path ($\mu$s-scale, Results~1--3), the DT for
what-if (${\sim}250$\,ms), and larger LLMs for explanation.
Open problems are calibrated uncertainty under distribution shift and
the maximum LLM size within a ${\leq}250$\,ms playbook budget.

\textit{Legacy OT, adversarial sensors and post-quantum crypto.}
IEC~60870-5-104, DNP3, Modbus and proprietary SCADA must be observed
non-intrusively via out-of-band taps and network slicing. A compromised PMU
or camera, or an FL poisoner, can mislead any single agent; defences
are cross-modality consistency checks (P1), DT plausibility filters
(P2) and signed anchor updates (P3). The anchor and CTI data
space must also outlive ``harvest now, decrypt later'' attacks,
motivating NIST post-quantum cryptography for ledger and STIX/TAXII transport.

\textit{Sovereignty and policy compliance.} CER/NIS2/AI-Act
conformance requires locally hosted open-weight LLMs, strict data
residency and explicit human-in-the-loop checkpoints on high-impact
actions~\cite{cisa2025_agentic,cisa2025_otai}; the challenge is to
codify these \emph{inside} the SOAR playbook generator rather than as
a post-hoc audit~\cite{cer2022,nis2}.

\section{Conclusion}\label{sec:conclusion}

RESILAGENT positions \emph{Sensing \& Data Intelligence},
\emph{Resilience Analytics \& Reporting} and \emph{Incident Response
\& Recovery} as three sides of one EPES resilience loop fused by
agentic AI, real-time DTs and a blockchain anchor. The contribution
is not another detection model but an \emph{operational architecture}
that integrates cyber, physical and environmental signals, drives a
twin-based decision surface, and produces verifiable CER/NIS2 evidence
by design, staged around four irreducible technologies
(Sec.~\ref{ssec:mvs}) with the rest as extensions. Result~6 runs the full
EWS$\to$DT$\to$detector$\to$vision$\to$LLM$\to$blockchain chain
end-to-end with a real Llama-3 agent, the decision path
completing in ${\sim}17$\,s, matched to the human-in-the-loop budget.
The evidence does not yet validate closed-loop performance on real
operator telemetry, imagery or a production ledger. A national
TSO and a parallel DSO pilot will close these gaps; the next steps
replace the three proxy components,
\textbf{(F1)} the detector and DT wrapped as tools inside a LangGraph
agent fed an LLM-parsed Suricata/Wazuh stream, \textbf{(F2)} a
multimodal LLM (LLaVA-class) benchmarked on Sentinel-2 and drone
imagery against a vision-blind baseline, and \textbf{(F3)} the
Open-Meteo demo replaced by the HYDS ARGOS multi-hazard EWS.

\section*{Acknowledgement}
This work has been partially supported by the CYBERGUARD project,
which has received funding from the European Union's Digital Europe
Programme under Grant Agreement No.~101190251.


\end{document}